\documentclass[apj]{emulateapj}

\bibpunct{(}{)}{;}{a}{}{,}

\slugcomment{in press, ApJL}
\shorttitle{HD 12661 Unexpected Evolution}
\shortauthors{Veras and Ford}

\begin{document}
\title{Secular Evolution of HD 12661: A System Caught at an Unlikely Time}
\author{Dimitri Veras\altaffilmark{1}, Eric B. Ford\altaffilmark{1}}
\altaffiltext{1}{Astronomy Department, University of Florida, 211 Bryant Space Sciences Center, Gainesville, FL 32111, USA}
\email{veras@astro.ufl.edu}
\begin{abstract}
The eccentricity evolution of multiple planet systems can
provide valuable constraints on planet formation models.
Unfortunately, the inevitable uncertainties in the current orbital
elements can lead to significant ambiguities in the nature of the
secular evolution.  Integrating any single set of orbital elements inadequately
describes the full range of secular evolutions
consistent with current observations.  Thus, we combine radial velocity
observations of HD 12661 with Markov Chain Monte Carlo sampling
to generate ensembles of initial conditions for direct
n-body integrations.  We find that any mean motion resonances 
are quite weak and do not significantly impact the 
secular evolution, and that current observations indicate 
circulation or large amplitude libration of the periapses.
The eccentricity of the outer planet
undergoes large oscillations for nearly all of the allowed
two-planet orbital solutions.  This type of secular evolution would
arise if planet c had been impulsively perturbed, perhaps due to
strong scattering of an additional planet that was subsequently
accreted onto the star.  Finally, we note that the secular evolution
implied by the current orbital configuration implies that planet c
spends $\sim~{96}\%$ of the time following an orbit more eccentric
than that presently observed.  Either this system is being observed
during a relatively rare state, or additional planets are affecting
the observed radial velocities and/or the system's secular
eccentricity evolution.
\end{abstract}


\section{Introduction}
The secular evolution of multi-planet systems has become a topic of
considerable importance for constraining planetary formation theories
\citep{foretal2005,adalau2006,bargre2006b,sanetal2007}.  We
investigate the secular evolution of the two giant planets orbiting HD
12661, a $\simeq~1.136 M_\odot$ star.  Planet HD 12661 b has a 
semimajor axis of $\simeq 0.83$AU and a
moderate eccentricity, and planet HD 12661 c 
follows a nearly circular orbit about three times
further away \citep{fisetal2003}.  This system has inspired several
dynamical studies.
A brief survey of their results demonstrates the importance of
considering the uncertainty in the current orbital configuration.
\cite{kisetal2002} considered coplanar edge-on systems and found that
the then-current best-fit solution was chaotic, behavior that is
likely due to close proximity to the 9:2 mean-motion resonance (MMR).
Both \cite{gozdziewski2003} and \cite{leepea2003} found that the
system was close to the 11:2 MMR and that the periapses underwent
large amplitude libration for the edge-on case, as well as for a broad
range of inclinations.  \cite{zhosun2003} claimed that aligned
configurations were less likely to be chaotic and thus more likely to
be stable.  \cite{gozdziewski2003} found the system could be chaotic,
but still stable for $\sim$Gyr thanks to a secular apsidal lock about
an antialigned configuration.  \cite{gozmac2003} performed an
independent analysis of the \cite{fisetal2003} radial velocities to
find a better orbital fit that placed the system very near (and likely
in) the 6:1 MMR that could be stabilized by secular apsidal lock for a
broad range of inclinations.  Despite the different orbital solutions,
both \cite{leepea2003} and \cite{gozmac2003} found that the periapses
were more likely to librate about an antialigned configuration, but
that librations about an aligned configuration were also possible,
particularly for systems with large inclinations.  However,
\cite{jietal2003} reported that aligned and antialigned configurations
were nearly equally likely.  Subsequently, \cite{butetal2006}
published an orbital solution with a ratio of orbital periods
approaching 13:2.
\cite{bargre2006a} found that this orbital solution places the 
system very near the the boundary between librating and 
circulating modes of secular evolution.
The revised orbital solutions for HD 12661 b \& c and the lack of
consensus regarding the system's secular evolution both motivate an
updated dynamical study.  Further, the historical range of orbital
configurations illustrates the importance of properly accounting for
uncertainties in orbital determinations.

\cite{gozdziewski2003} found that the classical secular theory gave
only a crude approximation to the secular evolution due to large
eccentricities and the 11:2 MMR.  Both \cite{verarm2007} and
\cite{libhen2007} caution against using low-order secular theories to
model the system's behavior.  \cite{leepea2003},
\cite{rodgal2005}, and \cite{libhen2007} found that either the
octupole or high-order Laplace-Lagrange approximation described the
secular evolution of HD 12661 well and that the secular dynamics is
not significantly affected by the 5:1, 11:2, and 6:1 MMRs.  Although
we do not expect near-resonant terms to significantly affect the
secular evolution, we use direct n-body integrations to 
account for the full dynamics, including all possible MMRs.

We determine the mode(s) of secular evolution
consistent with observations and the requirement of long-term
stability.  Our study improves upon previous studies by combining a
Bayesian analysis of an updated set of radial velocity observations
with direct n-body integrations to examine HD 12661's stability and
secular evolution.  We characterize the full range of orbital
histories that are consistent with observations and reject solutions
that do not exhibit long-term orbital stability.  Thus, we can make
quantitative statements about the relative probability of different
modes of secular evolution.

\section{Methods}

We reanalyze an updated set of radial velocity observations
(Wright et al.\ 2008) assuming that the stellar velocity is 
the superposition of two
Keplerian orbit plus noise.  First, we perform a brute force search
over parameter space to verify that there are no qualitatively different and
comparably likely solutions.  Working in a Bayesian framework, we
generate an ensemble of $\simeq~5\times10^5$ orbital solutions using
Markov Chain Monte Carlo (MCMC;
\citealt{ford2005,ford2006,gregory2007a,gregory2007b}).  Each state of
the Markov chain includes the orbital period ($P$), velocity amplitude
($K$), eccentricity ($e$), argument of pericenter measured from the
plane of the sky ($\omega$), and mean anomaly at a given epoch ($u$)
for each planet.  We use the priors, candidate transition functions,
automated step size control, and convergence tests described in
\cite{ford2006}.  In particular, we assume a prior for the velocity ``jitter'' ($\sigma_j$) of $p(\sigma_j) \propto [1+\sigma_j/(1 {\mathrm m s^{-1}})]^{-1}$.

We randomly select subsamples of orbital solutions to
be investigated with long-term n-body integrations and consider a
range of possible inclinations ($i$) and ascending nodes ($\Omega$).
We generate each orbit from an isotropic distribution (i.e., uniform
in $\cos{i}$ and $\Omega$) and divide the resulting systems into bins
based on the relative inclination between the two orbital planes
($i_{\rm rel}$).  The bins are
1) $i_{\rm rel}=0^{\circ}$ (coplanar), 
2) $0^{\circ} \le i_{\rm rel} \le 30^{\circ}$, 
3) $30^{\circ} \le i_{\rm rel} \le 60^{\circ}$, 
4) $60^{\circ} \le i_{\rm rel} \le 90^{\circ}$,
5) $90^{\circ} \le i_{\rm rel} \le 120^{\circ}$,
6) $120^{\circ} \le i_{\rm rel} \le 150^{\circ}$, and
7) $150^{\circ} \le i_{\rm rel} \le 180^{\circ}$.  
We integrate 2,000 systems in each of the first four bins and 500 in
each of the last three bins.  We can reweight our simulations so as to
determine the probability of the different modes of secular evolution
for various assumed distributions of relative inclinations.
From each set ($P,K,e,\omega,i,\Omega,u$), we generate the planet mass
($m$) and semi-major axis ($a$) using a Jacobi coordinate system
\citep{leepea2003}.

We used the hybrid symplectic integrator of {\tt Mercury}
\citep{chambers1999} to integrate each set of initial conditions for
at least 1 Myr.  Based on a smaller series of 10 Myr integrations, we
found that the vast majority of instabilities are manifest within 1
Myr.  We classified systems as ``unstable'' if, for either planet,
$a_{\rm max}-a_{\rm min} > \tau a_i$, where $a_{\rm max}$ and $a_{\rm
min}$ represent the maximum and minimum values of the semimajor axis,
$a_i$ is the initial value of the semimajor axis, and $\tau=0.3$.  We
discarded each set of initial conditions that was found to be
unstable, and analyzed the properties of the remaining ``stable''
systems.  Although a small fraction of our ``stable'' systems might
exhibit instability if integrated for much longer timescales, our
criteria avoids miscategorizing a system exhibiting bounded chaos
(e.g., \citealt{gozdziewski2003}) as unstable.  We manually verified
that the above criteria gives reasonable results and that various
choices of $\tau\in[0.10,0.30]$ make just a few percent difference in
the number of simulations labeled as stable.  The percent of stable
systems in our 7 bins are: 1) 100\%, 2) 99.7\%, 3) 93.3\%, 4)
16.4\%, 5) 0.2\%, 6) 16.7\%, and 7) 98.6\%. We independently affirmed
the trend exhibited by these stability percentages by performing a
smaller, additional set of simulations by using an ensemble of initial
conditions generated from a different MCMC code that accounted for planet-planet 
interactions with a Hermite integrator.

\section{Results}

Several previous studies of HD 12661
(e.g. \citealt{jietal2003,leepea2003,zhosun2003}) and other planetary
systems \citep{chietal2001,malhotra2002,foretal2005,bargre2006b} have
highlighted the importance of the apsidal angle
($\Delta\varpi=\Omega_b+\omega_b-\Omega_c-\omega_c$), which is useful
for describing the secular dynamics of systems with small relative
inclinations.
Since we consider systems with a wide range of relative inclinations,
we instead focus on $\Delta\varpi'$, the angle between the periastron 
directions projected onto the invariable plane.  Typically, the
apsidal angle is classified as circulating, librating about $0^\circ$,
or librating about $180^\circ$.  By inspecting plots of apsidal angle
evolution for individual systems, we find that some systems spend most
of the time librating about one center, but occasionally the apsidal
angle circulates for a short period of time.
Therefore, we consider two summary statistics describing the secular
evolution: the root mean square (RMS) of $\Delta\varpi'$ and the mean
absolute deviation (MAD) of $\Delta\varpi'$ about the libration
center, where the ``libration center'' is defined as the angle about which
the RMS $\Delta\varpi'$ is minimized.
%
%
For
a system undergoing small amplitude libration about either center, the
MAD and RMS deviation are of order the libration amplitude.  For a
system undergoing uniform circulation, the MAD approaches 90$^{\circ}$ and the
RMS deviation approaches 103.92$^{\circ}$.  These
statistics can be used to identify the mode of secular 
evolution in clear cut cases and provide a quantitative measure that is well-defined even for
systems with complex evolution.

We calculate both measures using orbital elements measured every $1000$ yr.  
Although the libration amplitude (maximum absolute
deviation of $\Delta\varpi'$) can be sensitive to the frequency of
sampling and leads to ambiguities, both RMS $\Delta\varpi'$ and MAD
$\Delta\varpi'$ are more robust statistics that allows us to describe
the secular evolution with a simple quantitative measure, regardless
of whether the system is librating, circulating, or switching between
regimes due to short-term perturbations.
%
%
This robustness is particularly important for studying systems where
there is little distinction between the librating and circulating
regimes, due to one planet's orbit periodically becoming nearly
circular.

For each n-body integration,
we determine the libration center and calculate both the RMS and MAD
of $\Delta\varpi'$ about that center.  
We discover that the RMS $\Delta\varpi'$ ranges from
$68^{\circ}-85^{\circ}$ and MAD $\Delta\varpi'$ ranges from
$63^{\circ}-82^{\circ}$, depending on the relative inclination
(see Table \ref{TabLib}).  The
libration centers are preferentially anti-aligned for
prograde, nearly-coplanar systems, and transition to 
almost entirely aligned for near-coplanar retrograde systems.  
The stability is highly dependent on initial relative
inclination; very few highly inclined retrograde systems were
stable, indicating that non-secular perturbations influence 
the long-term dynamics for some relative inclinations.  
Although large values of RMS $\Delta\varpi'$ and MAD $\Delta\varpi'$ 
may be consistent with non-uniform circulation, 
visual spot checks of individual systems all suggest libration.
%
%

%

%

If a system of two planets on nearly circular orbits is excited
impulsively, then the secular evolution will follow the boundary
between the circulating and librating regimes and cause one orbit to
repeatedly return to a nearly circular orbit \citep{malhotra2002}.  In
some systems, this secular evolution can be used to constrain the
system's formation \citep{foretal2005}.  \cite{bargre2006b}
found that the \cite{butetal2006} orbital solution implies the
eccentricity of HD 12661 c comes particularly close to zero, as
parameterized by their $\epsilon$ parameter (Eq. 1 of
\citealt{bargre2006b}).  Because they considered only the published
orbital solution in an edge-on, coplanar orientation, they were unable
to asses the finding's robustness to uncertainties in the orbital
parameters or non-edge-on systems.
In order to determine what fraction of stable systems consistent with
observations result in one planet returning to a nearly circular
orbit, we calculate four statistics from each of our n-body
integrations.  The first two are $c_1 \equiv e_{min,c}/e_{max,c}$ and
$c_2 \equiv e_{min,b}/e_{max,b}$, which represent the ratio of the
minimum to maximum eccentricity for each planet.  The 
third ($c_3$) is equal to 
$\left[ 2 {\rm min} (e_b e_c) \right]/
(x_{max} - x_{min} + y_{max} - y_{min})$, as defined by
\cite{bargre2006b}, where $x \equiv e_b e_c
\sin\left(\Delta\varpi\right)$ and $y \equiv e_b e_c
\cos\left(\Delta\varpi\right)$.
The fourth is 
$c_4 \equiv {\rm min}(e_b e_c) / 
            {\rm max}(e_b e_c)$,
which is a similar measure that includes the
eccentricities of both planets in a
rotationally symmetric manner that
is independent of the systems' orientation (unlike $c_3$).

\begin{figure}
\epsscale{0.7}
\plotone{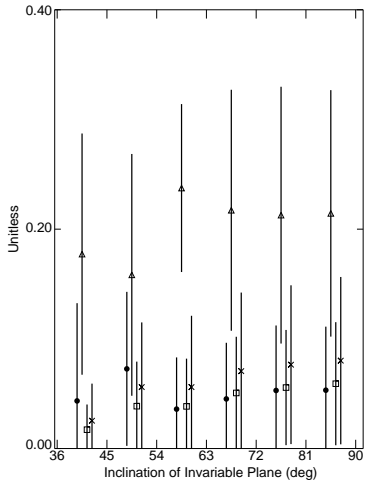}
\figcaption{Values of $c_1$ (dots), $c_2$ (triangles), $c_3$
(squares), and $c_4$ (crosses) and their standard deviations,
for coplanar systems, binned according to the sine of the inclination of the invariable plane. 
\label{FigCvsIlos}
\label{schem2}}
\end{figure}

\begin{figure}
\epsscale{0.7}
\plotone{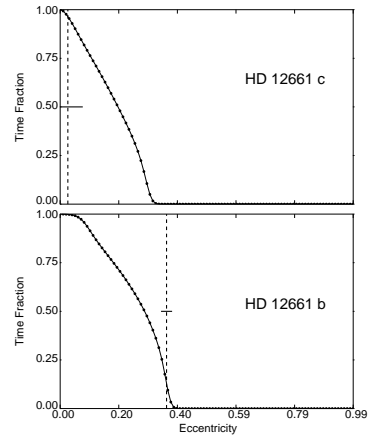}
\figcaption{Fraction of time for which each planet's eccentricity exceeded 
a given threshold (x-axis) averaged over allowed orbital solutions.  
The curves with dots show results for the coplanar case, but assuming an 
isotropic distribution of inclinations relative to the plane of the sky.  The dashed vertical 
lines indicate the median eccentricity values (0.031 and 0.361) for the planets' 
initial eccentricities, and their 5\%-95\% uncertainty ranges are
indicated by the horizontal solid segments. HD 12661 c's orbital eccentricity
exceeds its current value for approximately 96\% of its evolution.
\label{FigEHisto}
\label{schem1}}
\end{figure}

Figure \ref{FigCvsIlos} plots the mean and standard deviation of each
measure ($c_1$ - dots, $c_2$ - triangles, $c_3$ - squares, $c_4$ -
crosses) for coplanar systems which have been binned according to the
sine of the inclination of the invariable plane.  
The values
$c_1$, $c_3$ and $c_4$ all maintain values $< 0.1$ for all
bins of relative inclination for prograde systems, and 
$< 0.15$ for retrograde systems.  The large values of
$c_2$ (triangles) imply that HD 12661 b maintains a moderately 
eccentric orbit throughout the planet's evolution.  
Regardless, the measure $c_4$
demonstrates that the system lies near the boundary of libration and
circulation.  For the vast majority of systems, $c_3$ is over one 
order of magnitude greater than 0.003 (as found by \citealt{bargre2006b}).

The finding that planet c undergoes large eccentricity oscillations is
remarkable since it currently has an eccentricity of only
$\simeq~0.02$.  We investigate this observation further by plotting the fraction
of simulation time that each planet's eccentricity exceeded a given
value for the coplanar systems (Fig.\
\ref{FigEHisto}).  The dotted vertical lines indicate representative
values for the planets' initial eccentricities.  
HD 12661 c's orbital eccentricity exceeds its current
value for approximately 96\% of its evolution.  Thus, the
near-circular orbit we now observe is a rare occurrence.

\section{Conclusion}
Several previous studies had suggested HD 12661 b \& c are in a
secular apsidal lock and are undergoing large amplitude librations.
However, previously, even the orbital period of HD 12661 c was poorly
determined.  Now that radial velocity observations span over two
orbital periods of HD 12661 c, the current orbital period of planet c is
well-constrained.  An exhaustive search for
libration of all possible 11:2, 6:1, and 13:2 resonant angles
found that in only a few percent of all simulations performed, large
amplitude libration lasts over several $10^5$
yr.  The current orbital elements of the two
giant planets imply that the system lies near the boundary of
circulating and librating regimes, regardless of the unknown orbital
inclinations.  This behavior causes the eccentricity of planet c to
spend most of its time with a significant eccentricity ($e_c\ge~0.2$).  By computing the projected apsidal angle on the invariable plane, we find that the RMS $\Delta\varpi'$ ranges from $68^{\circ}-85^{\circ}$ and MAD $\Delta\varpi'$ ranges from $63^{\circ}- 82^{\circ}$, for any stable initial relative inclination, 
and planet c's eccentricity would be
greater than its current value $96\%$ of the time.

The most straightforward interpretation is that the eccentricity of
HD 12661 b was excited by an impulsive perturbation and planet c's
eccentricity is periodically excited via secular perturbations,
similar to the history of the $\upsilon$ And c \& d system
\citep{malhotra2002,foretal2005}.  Such an impulse could be the result
of planet-planet scattering involving at least one additional planet
\citep[e.g.][]{rasfor1996,weimar1996}.  The
current eccentricity of planet b suggests that the putative scattered
planet would have had a mass roughly half that of planet b 
\citep{forras2008}.  If the additional planet had been scattered outwards,
then there is a significant chance that it would have altered the
secular evolution of planet c \citep{bargre2006b}.  Given the
semi-major axis of planet b, it is possible that the scattered planet
was accreted onto the star.  If the instability occurred after the star
had reached the main sequence, then this instability could contribute to the
star's large surface high metalicity.  One potential concern for this
mechanism is whether the timescale for the apoastron of the scattered
planet to be lowered would be significantly shorter than the
$\simeq~10^4-10^5 $yr timescale for secular
evolution of planets b \& c.  

There is an alternative interpretation of our results.  One might 
think that we are unlikely to observe a system at such a rare time,
casting doubt upon the orbital fit and thus the inferred secular evolution.
One would expect that one of the 30 currently observed
multiple planet systems is transiently passing through such
an unusual state.  However, there are several ways in which
a system could be ``unusual'' so that the standard caveats of
{\em a posteriori} statistics apply.
Given the current radial velocity observations, there
are few alternatives.  We consider it unlikely that our global search
missed a qualitatively different pair of orbits due to our assumption
of non-interacting Keplerian orbits.  More plausibly, one or more
undetected planets could be affecting the orbital solution and/or the
secular evolution of the system.  After accounting for planets b \& c,
we find no statistically significant periodicities in the current
radial velocity observations.  Current observations are consistent
with a small long-term acceleration, but zero acceleration falls
within the 95\% credible interval.  Further, models with a long-term
acceleration do not result in significantly different orbital
parameters.  Our Bayesian analysis assuming two planets and 
uncorrelated Gaussian noise constrains the jitter to be
$\simeq~3.4\pm~0.7$m/s.  Further radial velocity observations can test
whether a portion of this jitter is due to additional planets.

\acknowledgments{We thank the referee for insightful comments, and
Debra Fischer and Geoff Marcy for providing
updated radial velocity observations prior to publication.  This
research was supported by NASA/JPL RSA1326409.  Some of the data
analyzed were obtained at the W. M. Keck Observatory, which is
operated as a scientific partnership among the California Institute of
Technology, the University of California, and the National Aeronautics
and Space Administration. The Observatory was made possible by the
generous financial support of the W. M. Keck Foundation.
We acknowledge the University of Florida High-Performance Computing
Center for providing computational resources and support.
}

\begin{deluxetable}{ c  c  c  c  c  c  c  c  c  c  c}
\label{TabLib}
\tabletypesize{\scriptsize}
\tablecaption{Results of N-body Integrations}
\tablewidth{0pt}
\tablehead{ 
   \colhead{$i_{rel}$} &
   \colhead{(RMS) aligned fraction} &
   \colhead{(MAD) aligned fraction} &
   \colhead{RMS $\Delta\varpi'$} &
   \colhead{MAD $\Delta\varpi'$} &
   \colhead{Stable} &
   \colhead{Isotropic} 
}
\renewcommand{\arraystretch}{1.5}
\startdata  
$0^{\circ}$   & $35.9\%$ & $31.7\%$ & $84.9^{\circ} \pm 8.5^{\circ}$ & $81.1^{\circ} \pm 7.8^{\circ}$ & $100\%$ &  \nodata
\\
$0^{\circ}-30^{\circ}$ & $39.3\%$ & $35.2\%$ & $83.8^{\circ} \pm 9.4^{\circ}$ & $79.5^{\circ} \pm 8.6^{\circ}$ & $99.7\%$ &  6.7\%
\\
$30^{\circ}-60^{\circ}$  & $56.8\%$ & $51.8\%$ & $84.2^{\circ} \pm 10.4^{\circ}$ & $79.7^{\circ} \pm 9.8^{\circ}$ & $93.3\%$ &  18.3\%
\\
$60^{\circ}-90^{\circ}$  & $56.0\%$ & $52.0\%$ & $83.8^{\circ} \pm 10.5^{\circ}$ & $79.6^{\circ} \pm 9.9^{\circ}$ & $16.4\%$ &  25.0\%
\\
$90^{\circ}-120^{\circ}$  & --- & --- & --- & --- & $0.2\%$ &  25.0\%
\\
$120^{\circ}-150^{\circ}$   & $95.2\%$ & $95.2\%$ & $74.7^{\circ} \pm 12.1^{\circ}$  & $69.3^{\circ} \pm 11.1^{\circ}$  & $16.7\%$ &  18.3\%
\\
$150^{\circ}-180^{\circ}$  & $99.6\%$ & $99.2\%$ &  $68.4^{\circ} \pm 12.8^{\circ}$ & $63.0^{\circ} \pm 12.1^{\circ}$ & $98.6\%$ &  6.7\%
\\ \hline
Prograde, isotropic  &  &  &  &  & 69.8\% & 80.0\% \\
Retrograde, isotropic  &  &  &    & &  38.5\% & 50.0\% \\
\enddata
%
\tablecomments{\label{TabLib} \small{
Columns 2 and 3 each list the RMS and MAD percent of stable systems with a libration center of $0^\circ$ instead of $180^\circ$ for the range of relative inclinations in column 1.  Columns 4 and 5 report the RMS and MAD variations for systems which librate around the most common center for that range of initial relative inclinations.  Column 6 lists the percent of stable systems, and Column 7 lists the percent of systems with the given range of inclination for an isotropic distribution.  The bottom two rows present aggregated stability statistics for the isotropic prograde and retrograde distributions of orbits.  2000 simulations were performed for each prograde relative inclination bin, and 500 simulations for each retrograde bin.}}
\end{deluxetable}


\begin{thebibliography}{}
%

\bibitem[Adams \& Laughlin(2006)]{adalau2006} 
Adams, F.~C., \& Laughlin, G.\ 2006, \apj, 649, 1004 

\bibitem[Barnes \& Greenberg(2006a)]{bargre2006a} 
Barnes, R., \& Greenberg, R.\ 2006a, \apjl, 647, L163 

\bibitem[Barnes \& Greenberg(2006b)]{bargre2006b} 
Barnes, R., \& Greenberg, R.\ 2006b, \apjl, 652, L53 

\bibitem[Butler et al.(2006)]{butetal2006} 
Butler, R.~P., et al.\ 2006, \apj, 646, 505 

\bibitem[Chambers(1999)]{chambers1999} 
Chambers, J.~E.\ 1999, \mnras, 304, 793

\bibitem[Chiang et al.(2001)]{chietal2001} 
Chiang, E.~I., Tabachnik, S., \& Tremaine, S.\ 2001, \aj, 122, 1607 

\bibitem[Fischer et al.(2003)]{fisetal2003} 
Fischer, D.~A., et al.\ 2003, \apj, 586, 1394 

\bibitem[Ford(2005)]{ford2005} 
Ford, E.~B.\ 2005, \aj, 129, 1706 

\bibitem[Ford(2006)]{ford2006} 
Ford, E.~B.\ 2006, \apj, 642, 505 

\bibitem[Ford et al.(2005)]{foretal2005} 
Ford, E.~B., Lystad, V., \& Rasio, F.~A.\ 
2005, \nat, 434, 873 

\bibitem[Ford \& Rasio(2008)]{forras2008} 
Ford, E.~B., \& Rasio, F.~A.\ 2008, 
\apj, 686, 621

\bibitem[Go{\'z}dziewski(2003)]{gozdziewski2003} 
Go{\'z}dziewski, K.\ 2003, \aap, 398, 1151 

\bibitem[Go{\'z}dziewski \& Maciejewski(2003)]{gozmac2003} 
Go{\'z}dziewski, K., \& Maciejewski, A.~J.\ 2003, \apjl, 586, L153 

\bibitem[Gregory(2007a)]{gregory2007a} Gregory, P.~C.\ 2007a, \mnras, 381, 1607 

\bibitem[Gregory(2007b)]{gregory2007b} Gregory, P.~C.\ 2007b, \mnras, 374, 1321 

\bibitem[Ji et al.(2003)]{jietal2003} 
Ji, J., Liu, L., Kinoshita, H., Zhou, J., 
Nakai, H., \& Li, G.\ 2003, \apjl, 591, L57 

\bibitem[Kiseleva-Eggleton et al.(2002)]{kisetal2002} 
Kiseleva-Eggleton, L., Bois, E., Rambaux, N., 
\& Dvorak, R.\ 2002, \apjl, 578, L145 

\bibitem[Lee \& Peale(2003)]{leepea2003} 
Lee, M.~H., \& Peale, S.~J.\ 2003, \apj, 592, 1201 

\bibitem[Libert \& Henrard(2007)]{libhen2007} 
Libert, A.-S., \& Henrard, J.\ 2007, \aap, 461, 759 

\bibitem[Malhotra(2002)]{malhotra2002} 
Malhotra, R.\ 2002, \apjl, 575, L33 

\bibitem[Rasio \& Ford(1996)]{rasfor1996} 
Rasio, F.~A., \& Ford, E.~B.\ 1996, Science, 274, 954

\bibitem[Rodr{\'{\i}}guez \& Gallardo(2005)]{rodgal2005} 
Rodr{\'{\i}}guez, A., \& Gallardo, T.\ 2005, \apj, 628, 1006

\bibitem[S{\'a}ndor et al.(2007)]{sanetal2007} 
S{\'a}ndor, Z., Kley, W., \& Klagyivik, P.\ 2007, \aap, 472, 981 

\bibitem[Veras \& Armitage(2007)]{verarm2007} 
Veras, D., \& Armitage, P.~J.\ 2007, \apj, 661, 1311 

\bibitem[Weidenschilling \& Marzari(1996)]{weimar1996} 
Weidenschilling, S.~J., \& Marzari, F.\ 1996, \nat, 384, 619 

\bibitem[Wright et al.(2008)]{wrightetal2008}
Wright, J.R., Upadhyay, S., Marcy, G.W., Fischer, D.A., Ford, E.B.\ 2008, \apj, submitted. 

\bibitem[Zhou \& Sun(2003)]{zhosun2003} 
Zhou, J.-L., \& Sun, Y.-S.\ 2003, \apj, 598, 1290


\end{thebibliography}
\end{document}